\documentclass[%
 aip,%
 amsmath,amssymb,
 preprint,%
 reprint,%
]{revtex4-1}

\usepackage{graphicx}
\usepackage{dcolumn}
\usepackage{bm}

\usepackage[utf8]{inputenc}
\usepackage[T1]{fontenc}
\usepackage{mathptmx}
\usepackage{etoolbox}

\usepackage{chemformula}
\usepackage{hyperref}

\makeatletter
\def\@email#1#2{%
 \endgroup
 \patchcmd{\titleblock@produce}
  {\frontmatter@RRAPformat}
  {\frontmatter@RRAPformat{\produce@RRAP{*#1\href{mailto:#2}{#2}}}\frontmatter@RRAPformat}
  {}{}
}%
\makeatother

\begin{document}

\title{Machine Learning Accelerates Raman Computations from Molecular Dynamics for Materials Science}
\author{David A. Egger}
\email{david.egger@tum.de}
\affiliation{Physics Department, TUM School of Natural Sciences, Technical University of Munich, 85748 Garching, Germany}
\affiliation{Atomistic Modeling Center, Munich Data Science Institute, Technical University of Munich, 85748 Garching, Germany}
\author{Manuel Grumet}
\affiliation{Physics Department, TUM School of Natural Sciences, Technical University of Munich, 85748 Garching, Germany}
\author{Tomáš Bučko}
\email{tomas.bucko@uniba.sk}
\affiliation{Department of Physical and Theoretical Chemistry, Faculty of Natural Sciences, Comenius University
in Bratislava, SK-84215 Bratislava, Slovakia}
\affiliation{Institute of Inorganic Chemistry, Slovak Academy of Sciences, SK-84236 Bratislava, Slovakia}

\date{\today}

\begin{abstract}
Raman spectroscopy is a powerful experimental technique for characterizing molecules and materials that is used in many laboratories.
First-principles theoretical calculations of Raman spectra are important because they elucidate the microscopic effects underlying Raman activity in these systems. 
These calculations are often performed using the canonical harmonic approximation which cannot capture certain thermal changes in the Raman response. 
Anharmonic vibrational effects were recently found to play crucial roles in several materials, which motivates theoretical treatments of the Raman effect beyond harmonic phonons.
While Raman spectroscopy from molecular dynamics (MD-Raman) is a well-established approach that includes anharmonic vibrations and further relevant thermal effects, MD-Raman computations were long considered to be computationally too expensive for practical materials computations.
In this perspective article, we highlight that recent advances in the context of machine learning have now dramatically accelerated  the involved computational tasks without sacrificing accuracy or predictive power.
These recent developments highlight the increasing importance of MD-Raman and related methods as versatile tools for theoretical prediction and characterization of molecules and materials.
\end{abstract}

\maketitle

\section{\label{sec:intro}Introduction}

The Raman effect has been discovered almost a century ago.
It is based on the inelastic scattering of light by vibrational excitations in substances such as molecules, liquids or crystalline materials.
The scattering of the electric field of the light by the atomic structure requires that a dipole moment is created when vibrational modes are excited. 
Vibrational modes can only create a dipole moment when they change the polarizability of the system, $\bm{\alpha}$. 
Vibrational modes that fulfill this criterion are called Raman-active modes.
This idea can be formalized using Raman selection rules, which depend on the symmetry of the vibrational modes: 
their symmetry dictates whether the atomic motions involved in vibrational modes can change $\bm{\alpha}$ or not.
An example that is relevant for crystalline systems -- in the strict limit of harmonic vibrational modes (so-called phonons) -- is that one can immediately predict which first-order Raman-active modes can exist in a given material from only knowing its crystalline symmetry. 

Raman spectroscopy is the analysis of the light that is inelastically scattered because of the Raman effect. 
This technique is widely applied in many experimental laboratories in order to provide detailed insight for vibrations in molecules and materials. 
A relatively recent development in the context of crystalline materials, which in part has been stimulated by Raman spectroscopy experiments, was the recognition of the effect that significantly anharmonic vibrations can have on the functional properties of materials. 
Vibrational anharmonicity describes a presence of higher-order terms in the potential energy surface of the crystal beyond the quadratic one that is encapsulated in the harmonic approximation.
When anharmonic vibrations occur, the thermally-activated motions of atoms will sample regions of the potential energy surface that deviate significantly from the parabolic shape of the harmonic approximation. 
Anharmonic vibrations are known to be closely associated with many thermal phenomena of crystals, such as thermal expansion, heat conduction and phase transitions.\cite{dove_lattice_dynamics}
However, more recently these anharmonic effects were found to strongly impact further key characteristics of materials including their optoelectronic properties, e.g., the fundamental band gap of various semiconductors.\cite{quarti_etal_2016,marronnier_etal_2018,zhou_bernardi_2019,zacharias_etal_2020,alvertis_engel_2022,gehrmann_etal_2022,kastl_etal_2023,schilcher_etal_2023,zacharias_etal_2023,seidl_etal_2023,hegner_etal_2024,zhu_egger_2025}

\begin{figure}
  \includegraphics[width=\columnwidth]{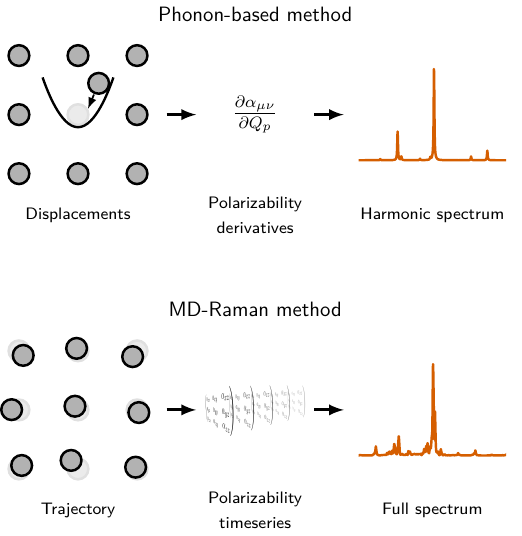}
  \caption{Conceptual overview showing the main steps of the phonon-based and MD-based methods for computing Raman specta.}
  \label{fig:sketch}
\end{figure}

Consequently, there is a strong demand to characterize anharmonic vibrations and their consequences for materials properties, and Raman spectroscopy appears to be an ideal tool for this task.
First-principles calculations of Raman spectra can offer a powerful synergy with experimental studies in order to characterize finite-temperature atomic motions in molecules and materials. 
Such calculations have long been established in the context of harmonic vibrations. 
For crystalline materials, the eigenvectors and dispersion relations of phonons can be calculated with electronic-structure methods such as density functional theory (DFT). 
Proceeding calculations using, e.g., density functional perturbation theory (DFPT), can then determine the change of $\bm{\alpha}$ for each of the DFT-calculated phonons. 
That is, these calculations determine the polarizability derivatives with respect to the phonon modes, $\frac{\partial {\alpha}_{\mu\nu}}{\partial Q_p}$, where $Q_p$ is the normal coordinate of a given phonon mode, $p$, and ${\alpha}_{\mu\nu}$ are the components of the $\bm{\alpha}$ tensor.
With the polarizability derivatives one can compute the Raman spectrum in the harmonic approximation, as illustrated in Fig.~\ref{fig:sketch}, top panel.

While this procedure is very well-established, several conceptual issues arise for calculations of essentially all finite-temperature properties when the considered chemical system has vibrations that are significantly anharmonic.
Focusing on Raman spectroscopy, we stress this point by highlighting the complete failure of predicting Raman-active modes based on crystalline symmetry for the case of cubic halide perovskites. 
While these materials should be Raman silent based on their average crystal symmetry, several studies showed that in fact they show significant Raman intensity, most prominently a broad feature at very low frequency that is known as "Raman central peak".\cite{yaffe_etal_2017,gao_etal_2021,huang_etal_2022,cohen_etal_2022,reuveni_etal_2023,lim_etal_2024,caicedo-davila_etal_2024}
Various experimental and computational studies have conclusively connected the notion of the Raman central peak in cubic halide perovskites to the appearance of strongly anharmonic vibrations in these systems,\cite{yaffe_etal_2017,menahem_etal_2023,gao_etal_2021,caicedo-davila_etal_2024,cohen_etal_2022} which in this specific case are anharmonic octahedral tilting motions.

Moreover and related to the above, when vibrational anharmonicity is strong and atoms sample higher-order regions of the potential energy surface, the average crystal structure might not coincide with the lowest-energy point on the surface, and in some cases a well-defined average structure might not exist at all.
These scenarios imply that the harmonic modes themselves will poorly approximate the actual atomic motions in the system that occur at finite temperature, especially when the average crystal structure is used in the Taylor expansion of the potential energy surface for calculating them.

The recent, significant interest in Raman spectroscopy of anharmonic materials motivates using alternative methods for first-principles computations of Raman spectra.
Fortunately, there exists a framework that is rooted in statistical mechanics and, in principle, treats anharmonic vibrations exactly. 
As will be explained in detail below, this statistical procedure is often realized using DFT-based molecular dynamics (MD) in conjunction with performing DFPT calculations along the trajectories for computing a polarizability timeseries, $\bm{\alpha}(t)$.

Fig.~\ref{fig:sketch}, bottom panel, illustrates this method in comparison to the standard phonon-based approach.
The benefit of the MD-based method, which we denote as MD-Raman, is that it can in principle be applied to any system and that it fully incorporates vibrational anharmonicity when calculating Raman spectra. 
In practice, however, the method is very expensive computationally as discussed in detail below.
Although the MD-Raman method and related methods for vibrational spectroscopy for molecular and materials science were developed several decades ago,\cite{berens_etal_1981,putrino_parrinello_2002,thomas_etal_2013,ditler_luber_2022} their computational burden has hindered their widespread application. 
This issue has motivated the development of alternative schemes that bypass MD calculations altogether, see, e.g., refs. \citenum{benshalom_etal_2022, knoop_etal_2024a, miotto_monacelli_2024,boziki_etal_2025}.

In this perspective article, we review the MD-Raman approach starting from the principles of statistical mechanics. 
This synopsis allows us to identify two major computational bottlenecks of MD-Raman that are associated with DFT-based MD and, most importantly, the required DFPT calculations for obtaining $\bm{\alpha}(t)$.
We highlight recent advances in the context of machine learning (ML) for rapid calculations of $\bm{\alpha}(t)$, which saw a tremendous progress in recent years. 
Already today, several methods exist that enable predictions of $\bm{\alpha}(t)$ fluctuations with \textit{ab-initio} accuracy at relatively low computational cost and can seamlessly be combined with  established and increasingly popular ML-based methods that accelerate DFT-based MD. We argue that these recent advances render ML-accelerated MD-Raman computations a versatile tool for theoretical characterization of molecules and materials.

\section{\label{sec:Raman_theory}The MD-Raman method}

\begin{figure*}
  \includegraphics[width=0.85\textwidth]{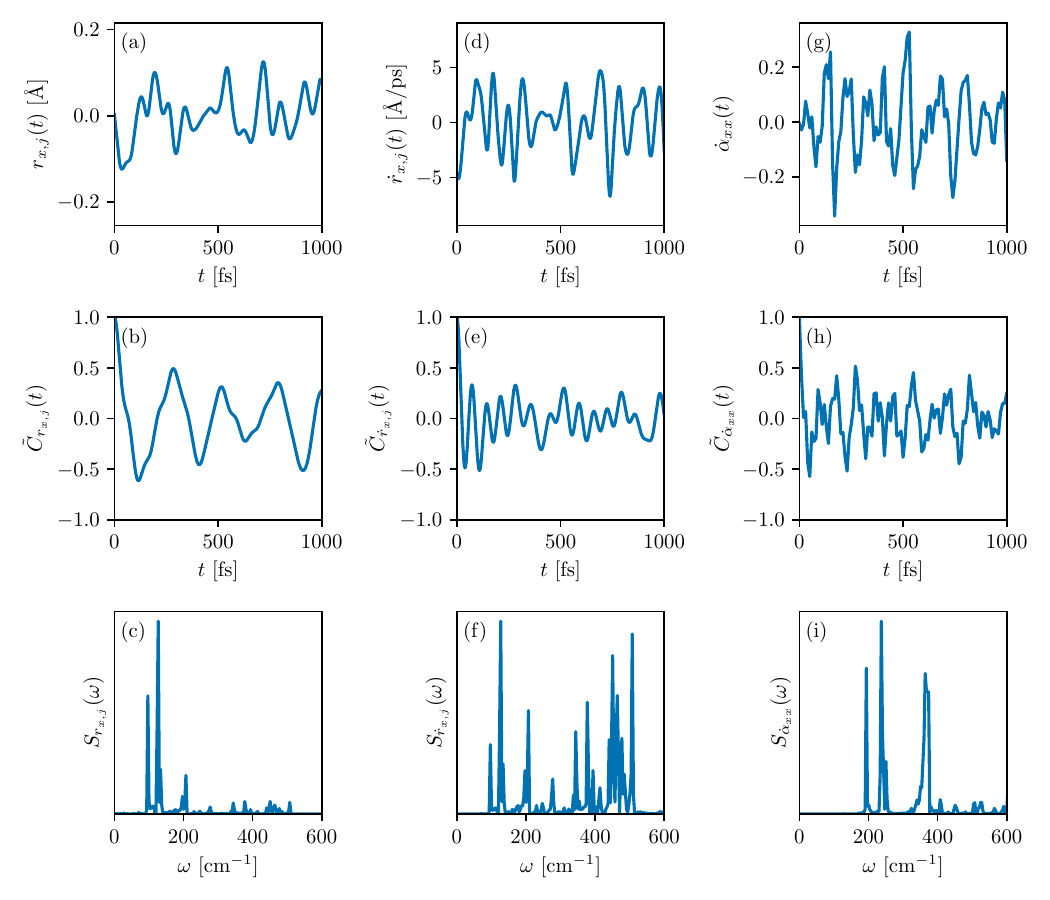}
  \caption{(a--c) Timeseries data, normalized autocorrelation function ($\tilde{C}_{\mathrm{r_{x,j}}}(t)=C_{\mathrm{r_{x,j}}}(t)/C_{\mathrm{r_{x,j}}}(0)$), and spectral density of the $x$ coordinate of a single atom in an Si crystal, obtained from an MD at simulation at a temperature of 300\,K. (d--f) Corresponding data for the $x$ component of velocity. (g--i) Corresponding data for the $xx$ tensorial component of the velocity of the dielectric tensor $\epsilon$, which is proportional to the polarizability tensor $\alpha$.}
  \label{fig:correlationfunctions}
\end{figure*}

One of the essential steps in the framework of the MD-Raman method is use of the spectral density and correlation function instead of normal modes. 
We use concepts from statistical mechanics to motivate this.~\cite{mcquarrie_statistical_mechanics}
Consider a continuous and stationary random process, $\chi(t)$.
Stationary means that the probability distributions will not change when time is shifted or, in other words, that the source of the random process does not depend on time.
For example, $\chi(t)$ may represent an $x$-component of the position vector of a particle $j$ ($r_{x,j}$)
experiencing Brownian motion as a function of time, $t$.
In Fig.~\ref{fig:correlationfunctions}a, we show an example for $r_{x,j}(t)$ as the position of an atom along one crystalline axis in a Si crystal, obtained from a DFT-based MD calculation at 300\,K that was reported in ref.~\citenum{grumet_etal_2024}.
An important object to describe such a process is the spectral density, $S(\omega)$, defined for a cartesian component of position coordinate, $r_{\mu,j}(t)$, as
\begin{equation}
 S_{\mathrm{r_{\mu,j}}}(\omega)=\lim_{T\to\infty}\frac{1}{2T}\lvert A_{\mathrm{r_{\mu,j}}}(\omega)\rvert^2,
\end{equation}
where $A_{\mathrm{r_{\mu,j}}}(\omega)$ is the Fourier transform of $r_{\mu,j}(t)$:
\begin{equation}
 A_{\mathrm{r_{\mu,j}}}(\omega)=\int_{-T}^T\mathrm{d}t\,r_{\mu,j}(t)\mathrm{e}^{-\mathrm{i}\omega t}.
\end{equation}
From the Wiener-Khinchin (WK) theorem for $\chi(t)$, i.e.,
\begin{equation}
 S_{\mathrm{\chi}}(\omega)=\int_{-\infty}^\infty\mathrm{d}t\,C_{\mathrm{\chi}}(t)\mathrm{e}^{\mathrm{i}\omega t},
 \label{eq:WK}
\end{equation}
we see that one can obtain the spectral density from Fourier transform of an autocorrelation function, defined for $r_{\mu,j}(t)$ as
\begin{align}
 C_{\mathrm{r_{\mu,j}}}(t)&=\lim_{T\to\infty}\frac{1}{2T}\int_{-T}^T\mathrm{d}\tau\,r_{\mu,j}(\tau)r_{\mu,j}(\tau+t)\nonumber\\
 &\equiv\langle {r_{\mu,j}}(\tau)\cdot{r_{\mu,j}}(\tau+t) \rangle_\tau.
 \end{align}
The correlation function is another central object for describing $\chi(t)$; we use a second time variable, $\tau$, when needed.
Again using the position of an atom in a Si crystal as an example, Fig.~\ref{fig:correlationfunctions}b and c show $C_{\mathrm{r_{x,j}}}(t)$ and $S_{\mathrm{r_{x,j}}}(\omega)$, respectively. 

First, we apply the formalism of correlation functions and spectral densities to analyze atomic motions in a physical system, i.e., without explicitly referring to the system's polarizability and the Raman effect. 
For this, we consider the velocity of a classical particle, $\dot{r}_{\mu,j}(t)$. 
While the frequency behavior of the velocity in principle contains the same information as the frequency behavior of the atomic position, choosing the velocity as the central quantity provides both conceptual and numerical advantages \cite{thomas_modeling_of_vibrational_spectra}.
Then, the per-particle velocity autocorrelation function, VACF, is defined as
\begin{equation}
    C_{\mathrm{\dot{r}_{\mu,j}}}(t) = \langle \dot{r}_{\mu,j}(\tau)\cdot\dot{r}_{\mu,j}(\tau+t) \rangle_\tau.
    \label{eq:Cv}
\end{equation}
From the WK theorem (Eq.~\ref{eq:WK}), we obtain the spectral density of the per-particle velocity as
\begin{equation}
 S_{\mathrm{\dot{r}_{\mu,j}}}(\omega)=\int_{-\infty}^\infty\mathrm{d}t\,C_{\mathrm{\dot{r}_{\mu,j}}}(t)\mathrm{e}^{\mathrm{i}\omega t}.
 \label{eq:SD_v}
\end{equation}
The entire spectral density of velocities, $\dot{\mathbf{r}}_{j}$, in a system of $N$ particles is obtained by summing over $S_{\mathrm{\dot{\mathbf{r}}}_j}(\omega)$, i.e.,
\begin{equation}
  S_\mathrm{\dot{\mathbf{r}}}(\omega)=\sum_{j=1}^N \sum_{\mu=x,y,z} S_{\mathrm{\dot{{r}}}_{\mu,j}}(\omega).
\end{equation}
Using Si as an example like above, we show MD-calculated $\dot{r}_{x,j}(t)$, $C_{\mathrm{\dot{r}_{x,j}}}$, and $S_{\mathrm{\dot{r}_{x,j}}}$ in Fig.~\ref{fig:correlationfunctions}d-f.

The vibrational density of states, VDOS, is an important quantity that can be obtained as mass-weighted sum of $S_{\mathrm{\dot{\mathbf{r}}}_i}(\omega)$, i.e.,
\begin{equation}
 g(\omega)=\sum_{j=1}^N m_j S_{\mathrm{\dot{\mathbf{r}}}_j}(\omega).
 \label{eq:VDOS}
\end{equation}
Note that we chose a definition without a prefactor here.
This means that in the purely harmonic case, $g(\omega)$ agrees with the phonon density of states only up to a frequency-independent factor.
From Eqs.~\ref{eq:Cv}, \ref{eq:SD_v} and \ref{eq:VDOS} we realize that $g(\omega)$ is related to the spectral density of the kinetic energy of the system. 
Hence, the VDOS provides information on the frequency-resolved contributions to the system's kinetic energy.

In order to achieve an analogous result for the Raman effect, we first stress that temporal changes of the polarizability, $\bm{\alpha}(t)$, are central for this process as described in the introduction. 
Hence, like for the particle velocities we define an autocorrelation function of polarizability velocities, PACF:
\begin{equation}
 C_{\dot{\alpha}_{\mu\nu}}(t) = \langle \dot{\alpha}_{\mu\nu}(\tau)\cdot\dot{\alpha}_{\mu\nu}(\tau+t) \rangle_\tau.
\end{equation}
Unlike velocity, however, $\bm{\alpha}$ and its time derivative are nonlocal quantities, i.e., one cannot provide atom-resolved contributions to it in a unique way.
However, one can analyze correlations in temporal changes of the tensorial components of $\bm{\alpha}(t)$, i.e., $\alpha_{\mu\nu}$ labeled with indices $\mu$ and $\nu$.
Using again the WK theorem (Eq.~\ref{eq:WK}), we obtain the spectral density of temporal changes of the components of $\bm{\alpha}$ as
\begin{equation}
 S_{\dot{\alpha}_{\mu\nu}}(\omega)=\int_{-\infty}^\infty\mathrm{d}t\,C_{\dot{\alpha}_{\mu\nu}}(t)\mathrm{e}^{\mathrm{i}\omega t}.
\end{equation}
This provides frequency-resolved contributions to the temporal changes of the system's polarizability, which is precisely what determines the magnitude of the Raman effect.
Fig.~\ref{fig:correlationfunctions}g-i show $\dot{\alpha}_{xx}(t)$, the $xx$-component of the temporal changes of $\bm{\alpha}(t)$, as well as  $C_{\dot{\alpha}_{xx}}(t)$, and $S_{\dot{\alpha}_{xx}}(\omega)$ for the MD-calculated example of bulk Si at 300\,K.

To reveal the connection between $S_{\dot{\alpha}_{\mu\nu}}(\omega)$ and the Raman effect, one can derive the Raman intensity for pure vibrational transitions \cite{long_the_raman_effect, placzek_1934}.
This derivation assumes the Born-Oppenheimer approximation, non-resonance, and purely vibrational transitions (so called Placzek vibrational transitions).
Furthermore, the expression for the Raman intensity $I(\omega)$ depends in principle on the orientation of the sample and on the illumination--observer geometry.
In the case of a spherical averaging,
one finds the following formula, which is valid
for a wide range of different illumination--observer geometries:
\begin{equation}
  I(\omega) \propto
  \frac{(\omega_\text{in} - \omega)^4}{\omega}
  \frac{1}{1 - \exp\left(-\frac{\hbar \omega}{k_\mathrm{B} T}\right)}
  \frac{45\,S_{a^2} + 7\,S_{\gamma^2}}{45}.
\end{equation}

Here, $\omega_\text{in}$ is the frequency of the laser beam, and $S_{a^2}$ and $S_{\gamma^2}$ are spectral densities characterizing temporal changes of tensor invariants $a^2$ and $\gamma^2$, that is the mean polarizability and anisotropy.
They can be defined for the symmetric $\bm{\alpha}(t)$ tensor and calculated from certain $S_{\dot{\alpha}_{\mu\nu}}$.
We note in passing that an analogous expression can be derived in the normal-mode framework, where $\frac{\partial {\alpha}_{\mu\nu}}{\partial Q_p}$ essentially takes the role of $S_{\dot{\alpha}_{\mu\nu}}$.

In summary, we have recapitulated the MD-Raman approach. 
It is a statistical framework for calculating Raman intensities from spectral densities of tensorial components of the polarizability velocity, i.e., $S_{\dot{\alpha}_{\mu\nu}}$.
It requires computing the effect of the atomic motions on the polarizability, i.e., the MD-Raman method necessitates calculations of $\bm{\alpha}(t)$ along an MD trajectory.

\section{\label{sec:bottleneck}Identifying the bottleneck of the MD-Raman approach}

\begin{figure*}
  \includegraphics[width=0.8\textwidth]{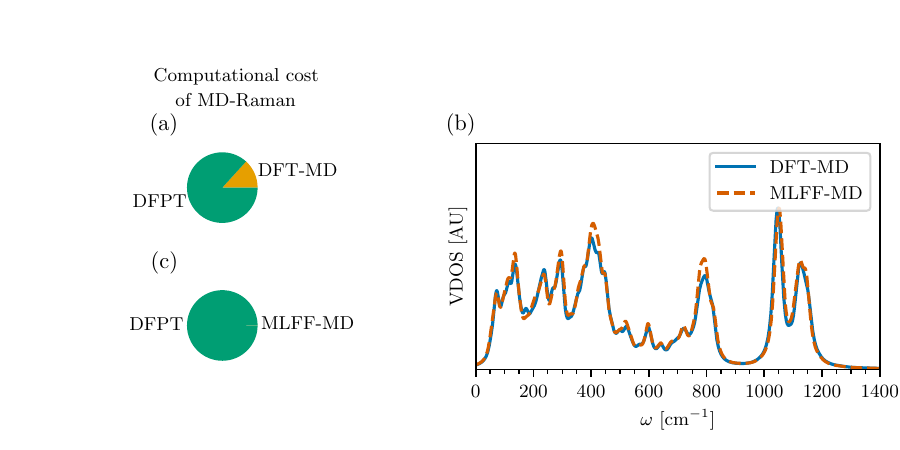}
  \caption{
  (a) Comparison of the relative computational cost of the MD simulation and the DFPT calculations in Raman-MD calculations without ML acceleration.
  (b) Comparison of the vibrational density of states (VDOS) of SiO$_2$ at 300\,K obtained using DFT-MD and MLFF-MD.
  (c) Comparison of the relative computational cost of the MD simulation and the DFPT calculations in Raman-MD calculations with MLFF-MD. 
  }
  \label{fig:SiO2_vdos}
\end{figure*}

We now discuss an example to analyze the computational costs involved in MD-Raman calculations from first-principles.
We focus on the crystalline material \ch{SiO2} as a test case and use data from ref.~\citenum{grumet_etal_2024}.
In brief, these data contain DFT-based MD calculations of a $3 \times 3 \times 3$ supercell of \ch{SiO2} at 300~K for 20~ps.

Fig.~\ref{fig:SiO2_vdos}a shows a comparison of the relative computational costs involved in calculating the Raman spectrum of \ch{SiO2} with MD-Raman.
While DFT-based MD calculations are generally computationally costly too, the major bottleneck in these simulations are the DFPT calculations for $\bm{\alpha}(t)$. 
In the example of \ch{SiO2}, approximately 85\% of computing time were spent on DFPT runs.

At this point, one may ask how many polarizability calculations are required for accurate predictions of $I(\omega)$. 
According to Nyquist's theorem, the maximum frequency that is resolvable in a spectrum is equal to half the sampling rate of the underlying timeseries.
Additionally, the number of data points that are obtained on the frequency axis after the Fourier transform is proportional to the number of data points in the timeseries.
Therefore, at a fixed sampling rate, the frequency resolution of the spectrum increases with the number of polarizability snapshots, i.e., with the length of the timeseries.

Furthermore, one needs to consider errors in the MD-calculated vibrational properties due to numerical approximations involved in Verlet integration.\cite{thomas_etal_2013, thomas_modeling_of_vibrational_spectra}
Their magnitude depends on the vibrational frequency and the MD timestep, with higher frequencies sampled using larger timesteps exhibiting greater error.
Considering a typical MD timestep of 1\,fs, the error can be significant -- approximately 40\,cm$^{-1}$ -- for a high vibrational frequency of 3000\,cm$^{-1}$.
But it becomes negligible, only 1\,cm$^{-1}$, for a vibration with a frequency of 1000\,cm$^{-1}$. 
Moreover, the numerical accuracy of the intensity in MD-predicted $I(\omega)$ is influenced by the sampling rate of $\bm{\alpha}(t)$, again depending on the vibrational frequency.
We have implemented corrections for both effects\cite{thomas_modeling_of_vibrational_spectra} but they do not affect the results we report here.
Altogether, several hundred up until around 1000 snapshots are required to accurately resolve all peaks in a Raman spectrum, which makes the MD-Raman approach computationally expensive.

Relevant in this context, the field recently witnessed an incredible development of ML force fields (MLFFs). 
Theoretical foundations of MLFFs and practical applications have been extensively discussed in several recent review articles.\cite{behler_2021,deringer_etal_2021,unke_etal_2021,batzner_etal_2023} 
Here, we only mention that while several topical issues such as incorporation of long-range effects remain, for many situations MLFFs can readily be used to obtain MD trajectories of molecules and materials with high accuracy compared to DFT.
The major advantage of MLFFs compared to purely DFT-based MD is the significant reduction of computational costs. 
In principle, these computational savings do not imply any compromise regarding calculation accuracy or empirical input to the calculation. 

To illustrate the impact of MLFFs in the context of MD-Raman calculations, we perform equivalent MD calculations for \ch{SiO2} as reported above, but now using the MLFF routines described in the Appendix.
Fig.~\ref{fig:SiO2_vdos}b shows a comparison of the VDOS, $g(\omega)$, for \ch{SiO2}, calculated with DFT and with a MLFF that has been trained using the same DFT approach. 
Overall, the agreement between VDOS calculated with DFT- and MLFF-based MD is very good, although minor deviations in the intensity of $g(\omega)$ are visible, particularly for lower-frequency modes.
Importantly, we stress that the overall computational costs to obtain $g(\omega)$ has been reduced by 98\% compared to the full DFT-based approach. 
Returning to the comparison of relative computational costs involved in MD-Raman, Fig.~\ref{fig:SiO2_vdos}c shows that the limiting effect of the DFPT calculations acting as bottleneck became dramatically more significant once DFT-based MD is replaced by a MLFF.
In case of \ch{SiO2}, approximately 98\% of computational costs of producing the Raman spectrum are now due to the DFPT runs to obtain $\bm{\alpha}(t)$.

The presented example underscores that predicting $\bm{\alpha}(t)$ is the computational bottleneck of MD-Raman. 
It is worth stressing that \ch{SiO2} is a relatively simple material in the sense that capturing the thermal motions of atoms around room temperature can be effectively achieved using moderately sized supercells containing several dozen to a few hundred atoms.
Since typical DFPT calculations scale as $N^3$, the bottleneck of performing the $\bm{\alpha}(t)$ calculations is more drastic in more complex systems that contain many atoms. 
Our example also illustrates that addressing the bottleneck issue in MD-Raman is key now that MLFF are becoming increasingly more popular in molecular and materials science. 
The original burden of performing DFT-based MD calculations has been massively diminished. 
With the increased availability of highly accurate MD trajectories of chemical systems, it becomes increasingly interesting to predict further quantum-mechanical observables. 
We will now discuss how this can be achieved for computing $\bm{\alpha}(t)$ in an ML-accelerated MD-Raman approach.

\section{\label{sec:Raman_ML}Tackling the bottleneck of the MD-Raman approach via machine learning}
\begin{figure*}
  \includegraphics[width=0.7\textwidth]{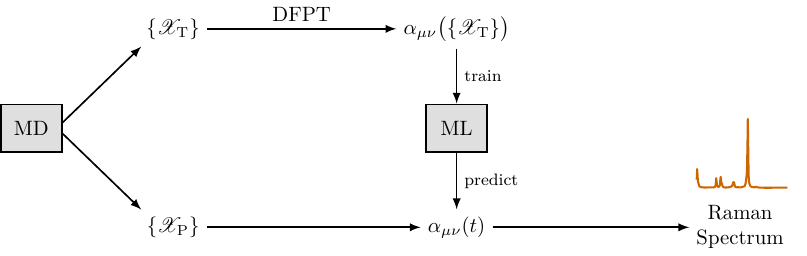}
  \caption{Conceptual overview of a MD-Raman-ML scheme. Snapshots from an MD trajectory are split into a training set $\{\mathcal{X}_\mathrm{T}\}$ and a prediction set $\{\mathcal{X}_\mathrm{P}\}$. Polarizabilities for the training data set are computed with DFPT. These are then used to train an ML model that can predict polarizabilities for the prediction dataset, thereby producing a complete timeseries $\alpha_{\mu\nu}(t)$.
  }
  \label{fig:ml_overview}
\end{figure*}

In the following, we will discuss ML methods for predictions of $\bm{\alpha}(t)$ in the context of Raman calculations. 
Next, we will demonstrate the power of these methods using the \ch{SiO2} example from above. 
Before discussing ML methods, we provide a brief historical account of polarizability models for chemical systems, again focusing on Raman spectroscopy.

Given a molecule or material, the simplest approach to computing the polarizability is to approximate it as a sum of atomic polarizabilities:
$\bm{\alpha}$ of the entire chemical system is approximated in a local representation of atomic polarizabilities. 
Since atomic polarizabilities are (approximately) isotropic, this naive model cannot capture important directionality properties of $\bm{\alpha}$.
Extending on this approach, Long and Bell proposed a bond polarizability model (BPM) for computing Raman intensities.\cite{long_bell_1952}
In the BPM, polarizabilities are assigned to bonds instead of atoms in the system, i.e., the method still uses a local representation and is strictly additive, but it can partly account for the inherent directionality of dielectric effects in chemical systems. 
Thole's model\cite{thole_1981,vanduijnen_swart_1998} is another approach that is also based on atomic polarizabilities but includes induction effects through assigning induced dipoles to atoms and self-consistently solving for $\bm{\alpha}$.

For several decades, these models and their derivatives have been widely applied in the context of Raman spectroscopy.\cite{montero_delrio_1976, long_the_raman_effect}
They have proven useful when interpreting Raman spectra of specific molecular\cite{snyder_1970,vanhemert_blom_1981,martin_montero_1984,barron_etal_1986,snoke_cardona_1993,guha_etal_1996} and condensed-phase systems,\cite{tubino_piseri_1975, zotov_etal_1999,umari_pasquarello_2002,smirnov_etal_2006,luo_etal_2015,bender_etal_2015,liang_etal_2017,dellostritto_etal_2019}
also when used together with first-principles calculations.\cite{umari_etal_2001,wirtz_etal_2005,hermet_etal_2006,paul_etal_2024}
In passing we mention recent studies that proposed alternative polarizability models, e.g., for Raman predictions of alloys and defective materials.\cite{hashemi_etal_2019, oleary_etal_2024} 
As DFPT-based MD-Raman calculations remain computationally intensive, the modeling of $\bm{\alpha}(t)$ with polarizability models such as the BPM along MD trajectories has emerged as a compelling alternative, drawing increasing interest in recent years.\cite{paul_etal_2024, paul_etal_2025, berger_komsa_2024, paul_grinberg_2025}

We now proceed with discussing ML methods for predicting $\bm{\alpha}(t)$. 
Fig.~\ref{fig:ml_overview} illustrates the conceptual idea behind the ML-accelerated MD-Raman method, which uses ML models for polarizability predictions in the MD-Raman approach. 
Our discussion focuses on ML methods that are trained on a system-per-system basis, but we will comment on methods that generalize across systems below where appropriate.
In passing, we note that ML models to predict Raman spectra without explicitly calculating $\bm{\alpha}(t)$ have also been proposed, see, e.g., refs. \citenum{hu_etal_2019, ren_etal_2021}.

We consider a task where for a specific chemical system, e.g., a molecule or crystalline material, a Raman spectrum at finite temperature needs to be calculated. 
To this end, we assume that a sufficiently long MD trajectory at the target temperature exists for the system at hand (see Fig.~\ref{fig:ml_overview}).
One then extracts snapshots for creating a training set from the MD data, i.e., one selects a set of atomic coordinates, $\{\mathcal{X}_\mathrm{T}\}$. 
Note that in principle it is also possible that the ML model itself selects the training data in an active learning scheme.\cite{gubaev_etal_2018}
DFPT calculations on these structures provide the polarizability training data, $\alpha_{\mu\nu}\left(\{\mathcal{X}_\mathrm{T}\}\right)$.
These training data are then used for a ML model that -- once trained sufficiently well -- delivers $\bm{\alpha}$ using only atomic coordinates as input. 
The ML model is applied to the prediction set, $\{\mathcal{X}_\mathrm{P}\}$, which provides $\bm{\alpha}(t)$ and delivers the Raman spectrum as in the MD-Raman formalism that we discussed above.

\begin{figure}
  \includegraphics[width=0.45\textwidth]{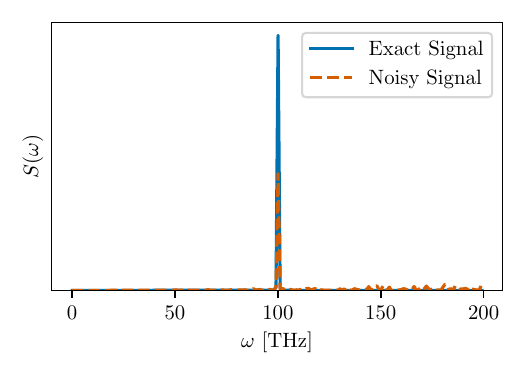}
  \caption{Demonstration of the effect of random noise on the spectral density, using synthetic data.
  The exact signal is a pure cosine with a single frequency of 100\,THz.
  The noisy version contains Gaussian noise that is independent and identically distributed each timestep.}
  \label{fig:noise}
\end{figure}

Prior to discussing different ML models, we mention a relevant conceptual issue, namely the influence of noise introduced by either ML-based $\bm{\alpha}(t)$ predictions or through other sources on the Raman spectrum, $I(\omega)$. 
To this end, we create synthetic signals (see Appendix), that is an exact signal in form of a pure cosine with a single frequency of 100\,THz and the same signal augmented by Gaussian noise that is independent and identically distributed each timestep, and compute $S(\omega)$ for both signals.
Fig.~\ref{fig:noise} shows that adding noise leads to reduction of the signal-to-noise ratio but does not affect the peak position.
Therefore, even though ML models will inevitably add noise to the $\bm{\alpha}(t)$ time series, this will not affect the predicted peak positions as long as the signal-to-noise ratio is high and the noise is statistically independent in time.
From this one can expect a certain tolerance of $I(\omega)$ against the presence of noise in the data.

The choice of ML model for $\bm{\alpha}$ in the approach sketched in Fig.~\ref{fig:ml_overview} is of key importance. 
Since such ML models have been recently reviewed in great detail \cite{han_etal_2022, ceriotti_2022, zhang_etal_2023a}, we focus here on highlighting selected key contributions. 
In doing so, we distinguish between two conceptually different families of ML models, namely kernel-based and neural-network-based schemes.

With kernel-based models, there exists a wide variety of descriptors that can be used for fitting polarizabilities.\cite{huang_vonlilienfeld_2021}
In 2018 Grisafi et al. introduced a technique to include rotational symmetry of tensorial properties in a Gaussian process regression (GPR) framework \cite{grisafi_etal_2018}. 
In this way, the smooth overlap of atomic positions (SOAP) kernel \cite{bartok_etal_2013} -- popular in the context of predicting scalar properties of atomistic systems with ML \cite{deringer_etal_2021, huang_vonlilienfeld_2021, musil_etal_2021, glielmo_etal_2021} -- was extended to the $\lambda$-SOAP method for learning tensorial quantities. 
Given the tensorial nature of $\bm{\alpha}$, in our view the development of $\lambda$-SOAP was an important step for ML predictions in the context of Raman spectroscopy. 
To the best of our knowledge, the first study adopting the approach shown in Fig.~\ref{fig:ml_overview} appeared shortly after $\lambda$-SOAP was proposed: 
Raimbault et al. predicted $\bm{\alpha}(t)$ with $\lambda$-SOAP to compute Raman spectra of molecular crystals.\cite{raimbault_etal_2019}
Comparing GPR frameworks with either SOAP or $\lambda$-SOAP, the authors observed significant improvement in model performance when tensorial learning was applied. 
They also proposed to include a baseline model for molecular crystals in GPR, which was built upon the molecular polarizability of non-interacting molecular units and found to further improve model performance.
In passing we note that in addition to these SOAP-based approaches, there also exist several other descriptors that can be used for kernel-based ML of polarizabilities, including the Coulomb matrix (CM), the bag of bonds representation (BoB), and the many-body tensor representation (MBTR).\cite{hansen_etal_2015}
These were successfully applied for Raman computations of organic molecules by Huo and Rupp. \cite{huo_rupp_2022}

Training a model to predict the difference between a fast, approximate method (the baseline model) and a more accurate, expensive one (in this case DFPT) is known as $\Delta$-ML. 
Building on the work of Grisafi et al. \cite{grisafi_etal_2018} and Raimbault et al. \cite{raimbault_etal_2019}, the present authors recently proposed a more general $\Delta$-ML model for polarizability predictions of molecules and crystals \cite{grumet_etal_2024}.
Using $\lambda$-SOAP as an ML technique, the baseline method consisted of a linear-response model approximating the dependence of $\bm{\alpha}$ on atomic displacements.  
Like in ref. \citenum{raimbault_etal_2019}, we found that a $\Delta$-ML approach can outperform a direct ML method and significantly reduce the training-set size required to reach accurate polarizability predictions.
For example, in the case of \ch{SiO2}, we found that the $\Delta$-ML approach required less than half the number of $\bm{\alpha}$ DFPT input values compared to direct ML.

The development of kernel-based ML tools for predicting $\bm{\alpha}$ is an ongoing and intense research effort.
For example, using coupled-cluster reference data and a $\lambda$-SOAP approach, Wilkins et al. proposed the AlphaML model for predicting polarizabilities across the chemical space of small organic molecules,\cite{wilkins_etal_2019} which was later applied to predict Raman spectra of alkanes.\cite{fang_etal_2024}
Related to this, Berger et al. trained kernel-based models on single amino acids to predict polarizabilities and MD-Raman spectra of peptides.\cite{berger_etal_2024}
An interesting alternative route to MD-Raman with kernel models starts from learning the static electronic density,\cite{lewis_etal_2021, rossi_etal_2025} from which the density response to external fields can be obtained for predicting Raman spectra of materials from MD.\cite{lewis_etal_2023}

Neural-network-based ML for predicting polarizabilities arguably have a longer history than kernel-based models, see, e.g., refs. \citenum{montavon_etal_2013,pronobis_etal_2018,sifain_etal_2018, nguyen_lunghi_2022, schutt_etal_2018}.
In the context of Raman spectroscopy, to the best of our knowledge Sommers et al. were the first to develop a deep neural network for learning $\bm{\alpha}$;\cite{sommers_etal_2020} note also ref. \citenum{zhang_etal_2020} that appeared shortly thereafter.
Starting from a local representation of $\bm{\alpha}$, in ref. \citenum{sommers_etal_2020} the symmetry and covariance properties of the tensor were addressed by combining two neural networks, one for embedding and one for fitting.
The authors utilized their ML approach for the polarizability together with a neural-network-based MLFF to compute Raman spectra of liquid water, modeling the temperature-dependence of different vibrational features across the spectrum.
MD-Raman calculations using deep neural networks predicting polarizabilities were quickly adopted, e.g., for highlighting anharmonic effects in vibrational spectra of crystalline materials.\cite{shang_wang_2021}
A crucial advancement enabling such predictive capabilities was developing a rotationally equivariant message passing framework within the so-called polarizable atom interaction neural network (PAINN) architecture \cite{schutt_etal_2021}.
This study demonstrated the importance of propagating directional information in message passing for MD-Raman calculations of molecules.\cite{gasteiger_etal_2020}

As is the case with kernel-based models, neural-network-based approaches for learning $\bm{\alpha}$ in the context of Raman spectroscopy are currently under active development.
This includes recent studies that proposed passing high-order tensors \cite{wang_etal_2024a}, high-dimensional neural network potentials \cite{chen_etal_2024}, neural networks that are designed to predict responses to external fields\cite{zhang_jiang_2023} or that are differentiable\cite{falletta_etal_2025}, and related methods such as neuroevolution potentials for tensorial properties.\cite{xu_etal_2024}

\begin{figure}
  \includegraphics[width=\columnwidth]{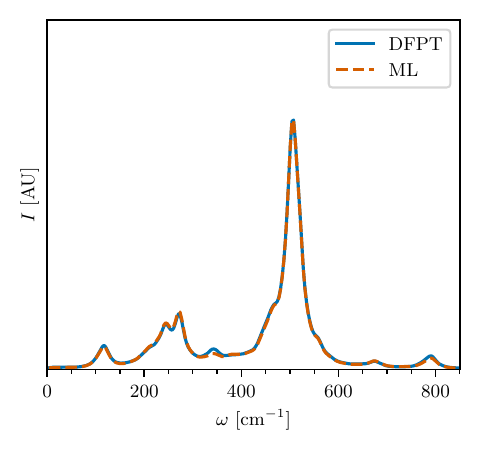}
  \caption{
  Comparison of Raman spectra of SiO$_2$ at 300\,K obtained using DFPT and Raman-ML, using the same underlying trajectory.
  Adapted from ref. \citenum{grumet_etal_2024} under the CC BY 4.0 license.
  }
  \label{fig:SiO2_spectra}
\end{figure}

The extensive developments of kernel- and neural-network-based models in the context of MD-Raman have been paralleled by various impressive applications. 
While no such list can ever be complete, we highlight predictions of hyper-Raman spectra,\cite{inoue_etal_2023} solvent effects on Raman spectra of molecules,\cite{gastegger_etal_2021} LO/TO splitting in materials, \cite{malenfant-thuot_etal_2024} and vibrational spectra of aqueous interfaces\cite{kapil_etal_2024} and bulk-phase chromophores.\cite{kebabsa_etal_2024}
Notably, these advances are increasingly supported by open-source software packages,\cite{schutt_etal_2023} which already today facilitate the integration of ML into Raman spectroscopy workflows.

We conclude this section by emphasizing the remarkable advances in ML approaches for polarizability predictions achieved in recent years. 
In Fig.~\ref{fig:SiO2_spectra} we show a comparison between MD-Raman calculations using DFTP- and ML-based $\bm{\alpha}(t)$ data for the example of \ch{SiO2}.
With the computational burden reduced by 95\%, and no visible reduction in accuracy for predicting a Raman spectrum, this example serves to highlight the impact of ML methods for applicability of the MD-Raman approach in molecular and materials science.

\section{\label{sec:conclusion}Conclusions and discussion}

We here discussed the statistical foundations and applications of MD-Raman, a computational method to predict Raman spectra of chemical systems without invoking the harmonic approximation and without relying on equilibrium structures of molecules or materials.
Because of these features, the MD-Raman approach is a powerful tool to investigate vibrational spectra at finite temperatures. 
Such capabilities are particularly useful in the context of studying chemical systems where anharmonic vibrational effects are important or where an equilibrium structure cannot be defined. 
The systems for which this is relevant span a wide range, from complex molecular systems such as proteins, to crystalline materials including organic and perovskite semiconductors or solid-state ion conductors, all the way to amorphous systems and liquids.
Although MD-Raman and related methods for vibrational spectroscopy have been proposed decades ago, the involved computational costs have so far limited their widespread use in the community.

Our synopsis on the elemental statistical principles of MD-Raman, together with an analysis of its computational costs for a test system, highlighted the major bottleneck involved in the approach: 
quantum-mechanical calculations of the required polarizability time-series, $\bm{\alpha}(t)$, amount to the lion’s share of the total simulation time -- often upwards of 80–90\% -- and they are the major source of computational costs. 
This is because one must perform a full DFPT calculation at many steps along the MD trajectory. 
As we have demonstrated, this is true already when \textit{ab-initio} calculations are used to generate MD trajectories that underlie MD-Raman. 
However, the issue becomes even more important when considering that today MLFFs can massively accelerate MD calculations of molecules and materials, exacerbating the bottleneck of predicting $\bm{\alpha}(t)$ as we have shown here. 

As a result, efficient and accurate methods for predicting $\bm{\alpha}(t)$ are in high demand to keep pace with these MD speed‐ups.
We started our discussion with polarizability models such as the BPM, which have long been applied in the context of Raman spectroscopy. 
We then reviewed the impressive progress of the past few years in developing ML models for predicting $\bm{\alpha}(t)$, where we distinguished between methods that use kernels and those based on neural networks. 
Both families of ML models heavily accelerate calculations of  $\bm{\alpha}(t)$ without compromising calculations accuracy and without the need for other input besides the computed data,
i.e., they operate fully from first-principles.
Interestingly, there is an analogy in the developments of simpler polarizability models and those concerned with ML methods. 
In both, including known physical principles in the model, e.g., accounting for the tensorial nature of $\bm{\alpha}$, has been found to improve performance and predictive power.
We expect that such physics-informed ML can positively influence future advances also in the domain of MD-Raman and in the context of predictions of quantum-mechanical observables more broadly.

The rapid
advances in ML models for predicting $\bm{\alpha}(t)$ that we have reviewed here eliminate the computational bottleneck involved in MD-Raman. 
Several ML methods for predicting polarizability fluctuations occurring along MD trajectories of chemical systems at low cost already exist now. 
Together with the impressive developments of MLFFs speeding-up MD runs, we conclude that MD-Raman and related methods for vibrational spectroscopy are emerging as versatile tools for molecular and materials science.

We now discuss interesting future directions in the context of ML methods for Raman predictions.
First, our results and discussions were based on a classical description of the nuclei, but it is important to stress that the here elaborated methods are not limited to it. 
For example, effects due to zero-point vibrations can be included in classical MD simulations, e.g., via a colored Langevin thermostat, and ML developments for full quantum treatments in MD are taking shape too, see, e.g., ref. \citenum{dammak_etal_2009}.
While the impact of coarse-graining $\bm{\alpha}(t)$ on the final Raman spectrum has been investigated above, we did not discuss the influence of the \textit{ab-initio} calculation itself. 
Clearly, any ML model for predicting $\bm{\alpha}(t)$ can at best only be as accurate as the ground-truth data it has been trained on. 
In this context, it is important to note that semilocal DFT functionals typically do not deliver accurate absolute polarizabilities, and that further numerical approximations involved in DFPT calculations of $\bm{\alpha}(t)$ can affect calculation accuracy too.
Hence, it will be important to benchmark how this affects the accuracy of ML models for Raman calculations that are trained with DFT data. 
Because there are several ML models available for $\bm{\alpha}(t)$ predictions as we have discussed above, systematic tests -- preferably conducted across diverse sets of molecules and materials -- will help determine faithful protocols for using MD-Raman in conjunction with ML.
Such efforts may well support key future developments, for example advancing representations for the nonlocal character of $\bm{\alpha}$ or further improving ML models for Raman predictions across chemical space.

\begin{acknowledgments}
The authors thank Frederico Delgado (TU Munich) for fruitful discussions. Funding provided by Germany's Excellence Strategy – EXC 2089/1-390776260 , and by TUM.solar in the context of the Bavarian Collaborative Research Project Solar Technologies Go Hybrid (SolTech), are gratefully acknowledged.
TB acknowledge support from the Slovak Research and Development Agency under the contracts No.
VEGA-1/0777/19 and
APVV-20-0127. 
The authors further acknowledge the Gauss Centre for Supercomputing e.V. for funding this project by providing computing time through the John von Neumann Institute for Computing on the GCS Supercomputer JUWELS at Jülich Supercomputing Centre.
\end{acknowledgments}

\section*{Data Availability Statement}

The data that support the findings of this study, including the MD-Raman tool and example calculations for \ch{SiO2}, are openly available on GitHub at \url{https://github.com/TheoFEM-TUM/MD-Raman}.

\appendix*
\section{Computational details}

As mentioned in the text, the data for DFT-based MD and DFPT of Si and \ch{SiO2} were taken from ref.~\citenum{grumet_etal_2024}.
In addition, we performed MLFF-based MD calculations for \ch{SiO2} using VASP \cite{kresse_furthmuller_1996, jinnouchi_etal_2019a}.
All numerical parameters related to DFT have been kept as described in ref.~\citenum{grumet_etal_2024}.
For MLFF training, starting from an already equilibrated configuration at 300\,K we perform an on-the-fly run where training configurations are selected based on Bayesian error estimation (BEEF) \cite{jinnouchi_etal_2019a}. 
Using a refit  of the available data based on singular value decomposition, we apply the trained MLFF in a production for calcualting the VDOS of \ch{SiO2}.
Furthermore, the synthetic data consist of a pure cosine wave with a single frequency of 100\,THz, which was discretized as a timeseries with a timestep of 1\,fs and a total length of 1000 points.
We added the noise independently at each timestep in form of a Gaussian distributed with $\sigma=1.0$, measured relative to the amplitude of the cosine.

\bibliography{references.bib}

\end{document}